\documentclass[aps,prd,reprint,preprintnumbers,superscriptaddress,nofootinbib,longbibliography,floatfix]{revtex4-2}

\usepackage{amsmath,amssymb,graphicx,bm,physics,slashed,bbm,multirow,comment}
\usepackage{hyperref}
\usepackage{orcidlink}
\usepackage{verbatim}

\begin{document}

\preprint{COMETA-2026-49}

\title{Light fermions can trigger a Brout-Englert-Higgs effect}

\author{Axel Maas\,\orcidlink{0000-0002-4621-2151}}
\email{axel.maas@uni-graz.at}
\affiliation{Institute of Physics, NAWI Graz, University of Graz, Universitätsplatz 5, 8010 Graz, Austria}

\author{Sofie Martins\,\orcidlink{0000-0001-5428-8925}}
\email{sofie.martins@uni-graz.at}
\affiliation{Institute of Physics, NAWI Graz, University of Graz, Universitätsplatz 5, 8010 Graz, Austria}

\author{Georg Wieland\,\orcidlink{0009-0003-1777-4563}}
\email{georg.wieland@uni-graz.at}
\affiliation{Institute of Physics, NAWI Graz, University of Graz, Universitätsplatz 5, 8010 Graz, Austria}

\date{\today}

\begin{abstract}
The Brout-Englert-Higgs effect produces the masses of the Standard Model particles. However, the cause of this effect is unknown. Beyond-the-Standard-Model theories aim to find the source of this effect, for example using a composite Higgs and dynamically generating its vacuum expectation value. In this paper, we show evidence from lattice simulations with vectorially coupled fermions and without Yukawa couplings that the Brout-Englert-Higgs effect could be triggered by the light masses of fermions. This suggests that an explanation entirely within the Standard Model could be possible. Moreover, our results are consistent with weak couplings and fermions can only impact the Higgs potential at NNLO. This implies a so far unidentified mechanism in an SU(2) scalar-fermion-gauge system that produces a non-zero vacuum expectation value. In addition, we find that the transition from a QCD-like behavior to a Higgs-like behavior happens when the fermion mass falls below twice the W/Z mass: a threshold that is integer valued but not connected to other critical physical behavior. 
\end{abstract}

\maketitle

{\bf Introduction} -- Identifying the origin of the Brout-Englert-Higgs~(BEH) effect is one of the central aims of modern particle physics~\cite{deBlas:2025gyz}. In the Standard Model, at tree-level, there appears to be no direct dynamical origin, but rather it is a consequence of particular properties of the Higgs potential~\cite{Bohm:2001yx,Maas:2017wzi}. In particular, the decisive sign of the quadratic term appears as an input. Nonetheless, this sign is sensitive to radiative corrections due to the Yukawa interactions of the Higgs with the fermions, as the discussion on the electroweak vacuum in recent years amply demonstrates~\cite{Sher:1988mj,Elias-Miro:2011sqh,Alekhin:2012py,Iacobellis:2016eof}. However, this is not without issues~\cite{Holland:2003jr,Branchina:2014rva}, especially beyond perturbation theory~\cite{Bulava:2012rb,Eichhorn:2015kea,Borchardt:2016xju}. Likewise, the fermions are one of the proposed mechanisms to link the breaking of supersymmetry to the appearance of a BEH effect, see~\cite{Aitchison:2007fn} for an introduction. This can take even more drastic forms when replacing the Higgs with new fermions altogether, see~\cite{Andersen:2011yj} for an example, or in (partial) compositeness scenarios~\cite{Kaplan:1991dc,Sannino:2016sfx}.

And then there are, of course, Abbott-Farhi-type models~\cite{Abbott:1981re}, linking the mechanism to effective non-perturbative dynamics above the electroweak scale. These usually also admit vectorial fermions, and thus are amendable for lattice simulations~\cite{Lee:1988ut,Hsu:1993zc,Lee:1987zu,Aoki:1988aw,Aoki:1988fg,Hansen:2017mrt}. There are similarities to other ideas of a non-perturbative, confining nature of the electroweak physics, rather than due to a perturbative BEH effect alone~\cite{Dosch:1983hr,Dosch:1984ec,Calmet:2000th,Calmet:2001rp,Calmet:2002mf}.

Here, we add another aspect to this cornucopia of possibilities. We study an SU(2)~Yang-Mills theory with two vectorially coupled fermions in the fundamental representation, and a fundamental Higgs doublet. This theory acts as a proxy to the Standard Model for lattice simulations of leptons and their electroweak interactions~\cite{Afferrante:2020fhd,Maas:2026dlq}, see also~\cite{Aoki:1988fg,Aoki:1988fg,Lee:1987zu,Lee:1988ut,Lee:1987eg,Greensite:2026jim}. This is, in principle, a technical tool, as chiral fermions cannot yet be simulated on the lattice~\cite{Hasenfratz:2007dp}. Aside from this minor technical detail, the results agree, at least for some range of the parameters, very well~\cite{Maas:2026dlq} with the expectations~\cite{Maas:2017wzi} for the dynamics in case of a BEH effect. In particular, the presence of the Higgs protects the dynamics from switching into a quasi-conformal phase when the fermions become light. The latter is otherwise observed~\cite{Andersen:2011yj}.

This is what happens when starting at infinite fermion mass from the bosonic theory when a BEH effect is happening~\cite{Maas:2026dlq} (called Higgs-like domain in the following). But without fermions, the theory has also regions of its phase diagram where no BEH effect takes place (called QCD-like domain in the following). In this case, no Higgs condensate can be observed in any gauge~\cite{Caudy:2007sf,Greensite:2008ss}, and the dynamics are that of a scalar version of QCD~\cite{Maas:2017wzi}. When adding fermions starting from a point in this region, the first expectation is that the theory keeps these features, see, e.g.,~\cite{Iida:2007qp} for a similar situation but for an SU(3)~gauge group.

At high fermion mass, we indeed will confirm these expectations. But when lowering the mass, we observe a different behavior. When the fermion mass is around twice the vector boson mass, the dynamics change and a BEH effect takes place. In fact, for all purposes, we observe the same dynamics~\cite{Maas:2026dlq} as if we would have started with a BEH effect at infinite fermion mass. This qualitative behavior persists down to the lowest masses, which were accessible to us, though there are slight differences compared to the former case. Nevertheless, the fermions becoming light triggers the BEH effect.

{\bf Theory and lattice methods} -- We study a proxy theory of the Standard Model by considering an SU(2)~scalar-fermion-gauge system with the scalar field in the fundamental representation and two mass-degenerate generations of vectorial leptons using the \texttt{HiRep} code base~\cite{DelDebbio:2008zf, Hansen:2017mrt, Drach:2025eyg}. For Wilson-Dirac fermions, the lattice parameters are given by the bare gauge coupling~$\beta$, the bare scalar mass~$(am_S)^2$, the quartic coupling~$\lambda$, and the fermion hopping parameter in the range ~$\kappa_F = 0.05-0.14$. Since we set the Yukawa couplings to zero, equivalent right-handed (ungauged) leptons decouple and we are not able to distinguish between leptons and neutrinos as their only difference is indicated by a gauge index.

Lattice simulations require gauge-invariant, physical observables, which are necessarily composite. Their properties are linked to the usual Higgs and vector bosons by the Fr\"ohlich-Morchio-Strocchi~(FMS) mechanism~\cite{Frohlich:1980gj,Frohlich:1981yi}. We refer to~\cite{Maas:2026dlq,Maas:2017wzi,Afferrante:2020fhd} for details. The masses of physical composite states can be mapped one-to-one to the elementary ones without the necessity of gauge fixing. We will measure three channels motivated by the FMS~mechanism: a scalar singlet~(Higgs boson), a vector triplet~(W/Z boson), and a fermion doublet~(weakly charged lepton). Furthermore, we extract the masses in four leptonium channels (pseudo-scalar, scalar, vector, and axial-vector) and determine the pseudo-scalar decay constant. The leptonia are described by the same operators one would use for mesons in QCD. To compute the vacuum expectation value, we fix to minimal 't~Hooft-Landau gauge. For a detailed discussion on the analysis of the lattice data, consult~\cite{Maas:2026dlq}.

{\bf Results} -- From simulations in the Higgs-like domain~\cite{Maas:2026dlq}, it was found that the fermion mass and quantities associated with it are the only observables of the system that experience significant changes when the fermion input parameter is altered. Our aim here is to test this fact starting from the QCD-like domain. Nevertheless, for presenting the data, we still assume that the fermion mass is the most relevant quantity and therefore, we plot all results against ratios of the fermion mass\footnote{Note that we use lattice units exclusively as scale setting cannot be performed uniformly for the different domains of the theory.}. The numerical values for all observables discussed in this section can be found in the dedicated data release~\cite{data_release}.

\begin{figure*}
    \centering
    \includegraphics[width=0.48\linewidth]{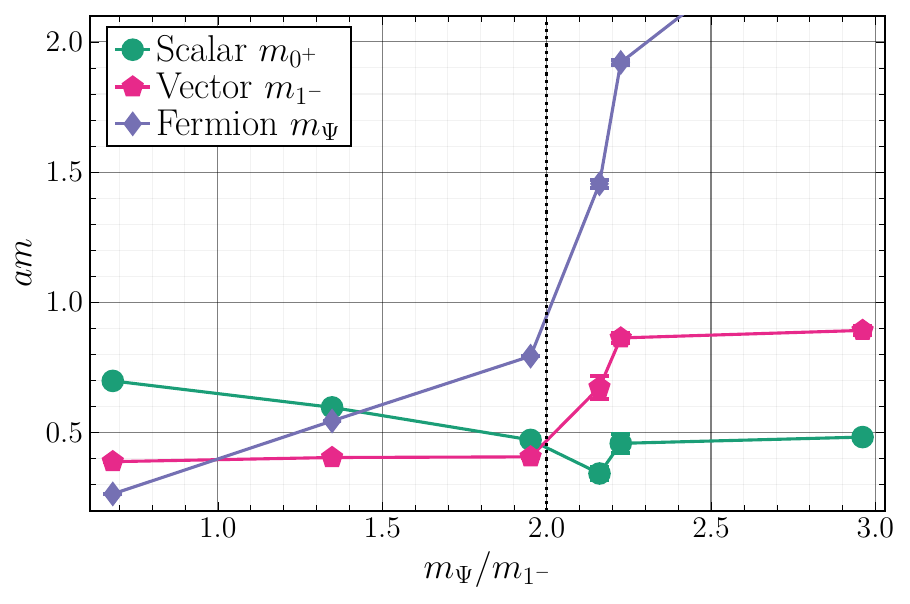} \hfill
    \includegraphics[width=0.468\linewidth]{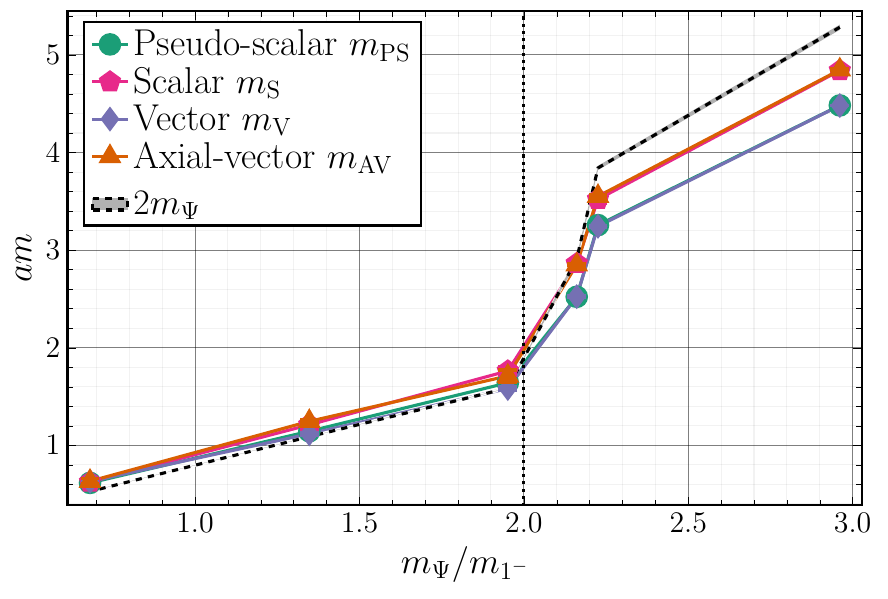}
    \caption{Left: Spectrum of the FMS operators in lattice units plotted against the ratio~$m_\Psi/m_{1^-}$. The change in mass hierarchy for the scalar and vector bosons happens around~$m_\Psi/m_{1^-} = 2$. Right: Ground state masses of all considered leptonium channels in lattice units against~$m_\Psi/m_{1^-}$. The plot shows that the transition from stable (below the $2m_\Psi$~threshold) to unstable (above the $2m_\Psi$~threshold) occurs around~$m_\Psi/m_{1^-} = 2$.}
    \label{fig:spec}
\end{figure*}

The left plot in Fig.~\ref{fig:spec} depicts the energy spectrum of the FMS observables, which represent the Higgs, W/Z boson, and lepton of the theory in the Higgs-like domain. Starting with heavy fermions, we find that the vector mass is almost twice the scalar mass. Reducing the fermion mass reveals deviations in both vector and scalar mass while keeping the hierarchy fixed. At some point around~$m_\Psi/m_{1^-} = 2$, the mass hierarchy flips and the vector triplet assumes an almost constant ground state mass for the next three data points. After this flip in the mass hierarchy, the scalar mass increases. This is the first indication for a genuine transition from the QCD-like to the Higgs-like domain of the system triggered by a reduction in fermion mass.

More evidence for this behavior can be found when looking at the masses of different leptonia in the theory, see the right plot in Fig.~\ref{fig:spec}. For all channels, a clear picture emerges: For heavy fermions (in the QCD-like domain), the ground state masses of all leptonium channels are below the threshold of two leptons. This makes all leptonia stable states as they cannot decay into two leptons. For lighter fermions (in the Higgs-like domain), all leptonia become consistent with being scattering states of two leptons. This is in line with simulations in the Higgs-like domain~\cite{Maas:2026dlq}. Note that in the QCD-like domain the masses of parity partners are not degenerate, a signal of chiral symmetry breaking. Nonetheless, the pseudo-scalar and the vector mesons are degenerate within our resolution. This implies that the theory is far from the chiral limit, as is expected from the relative sizes of the leptonia masses compared to the other mass scales.

\begin{figure*}
    \centering
    \includegraphics[width=0.48\linewidth]{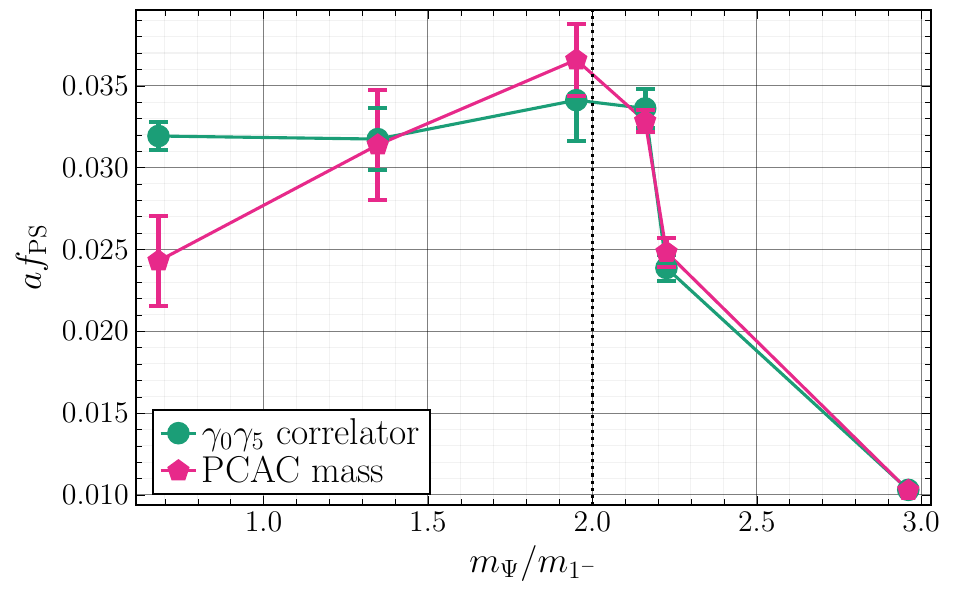} \hfill
    \includegraphics[width=0.464\linewidth]{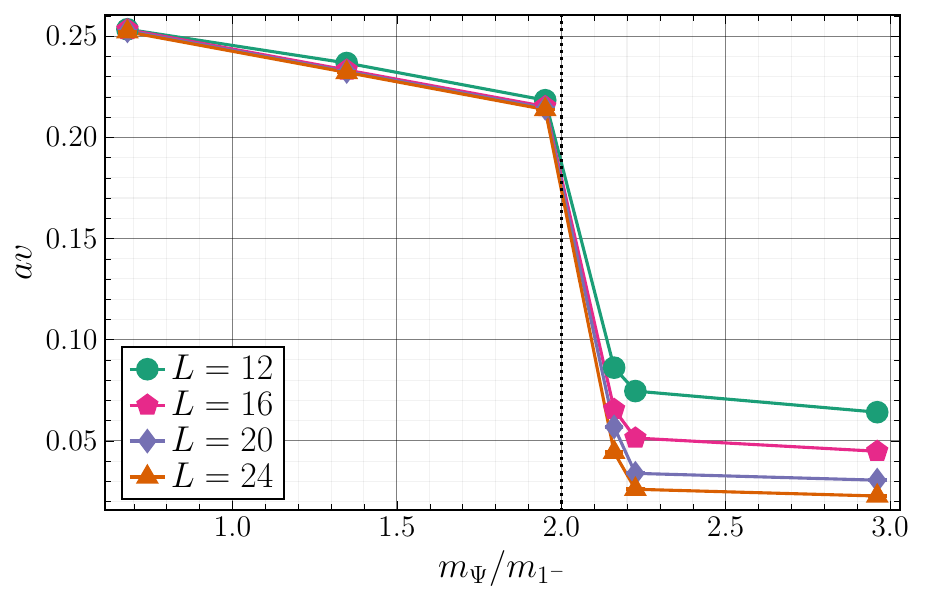}
    \caption{Left: Pseudo-scalar decay constant as a function of~$m_\Psi/m_{1^-}$ in lattice units. The decay constant is extracted in two ways: once from the $\gamma_0\gamma_5$~correlator and once from the PCAC mass. See~\cite{Maas:2026dlq} for details. Right: Vacuum expectation value~$av$ for different lattice sizes~$L$. There is a clear transition from a volume independent behavior for~$m_\Psi/m_{1^-} \lesssim 2$ to a volume dependent behavior for~$m_\Psi/m_{1^-} \gtrsim 2$. For~$m_\Psi/m_{1^-} \lesssim 2$, we find a non-vanishing vacuum expectation value, indicating a transition into the Higgs-like domain of the theory.}
    \label{fig:add}
\end{figure*}

Besides the spectroscopy observables discussed above, we also extracted the renormalized pseudo-scalar decay constant~$af_{\text{PS}}$ using two different methods, see the appendix in~\cite{Maas:2026dlq} for details. The left plot in Fig.~\ref{fig:add} depicts the results for both methods. Here, we would like to highlight two key findings: First, the two methods agree within errors in the QCD-like domain but differ in the Higgs-like domain. This is expected, as it is assumed that chiral perturbation theory applies in the QCD-like region, which in turn implies that the method of extracting the pseudo-scalar decay constant using the PCAC mass is valid and should therefore agree with the alternative method. Second, we observe an increase of~$af_\text{PS}$ towards the transition into the Higgs-like domain. Note, however, that the last data point in the QCD-like domain is of the same magnitude in lattice units.

As the mass in the $1^-$~channel stabilizes in the Higgs-like domain, we are able to use this quantity to fix to physical units by requiring $m_{W/Z} = 80.375$~GeV. This allows us to estimate the pseudo-scalar decay constant in the Higgs-like domain in physical units. A comparison with the results of the simulations in~\cite{Maas:2026dlq} reveals that for all ensembles in the Higgs-like region,~$f_{\text{PS}}$ is of the same order of magnitude, namely of~$\mathcal{O}(10)$~GeV.

For the vacuum expectation value, shown in the right plot in Fig.~\ref{fig:add}, we find a scaling with the lattice extent in the QCD-like domain, whereas its volume dependence is dominated by a constant contribution in the Higgs-like region. This gives a finite vacuum expectation value for~$m_\Psi / m_{1^-} \lesssim 2$, indicating an active BEH effect. In the QCD-like domain, the scaling indicates a vanishing vacuum expectation value for~$L \rightarrow \infty$.

{\bf Behavior of the observed transition} -- The results we find delineate two distinctively different regions. The one is the region~$m_\Psi/m_{1^-} \gtrsim 2$. Here, we find signs of leptonia as bound states, indications for dynamical chiral symmetry breaking, and no sign of a Higgs vacuum expectation value in a fixed gauge. In addition, we find~$m_{0^+}/m_{1^-}<1$. All features are consistent with expectations for QCD-like dynamics, including the existence of a state, which can be identified as a chimera particle~\cite{Iida:2007qp}, the Higgs-lepton bound state.

For~$m_\Psi/m_{1^-} \lesssim 2$, we do not find any bound leptonia, a non-vanishing Higgs vacuum expectation value, no indications of dynamical chiral symmetry breaking, and signs of the FMS mechanism. In addition, we find~$m_{0^+}/m_{1^-}>1$. The absence of bound leptonia disfavors an interpretation in terms of partial compositeness. In fact, almost all evidence is consistent with the same type of Higgs-like domain as observed in other regions of the phase diagram~\cite{Maas:2026dlq}. The only difference is that the scalar/Higgs mass changes when decreasing the ratio~$m_\Psi/m_{1^-}$ further. This translates, as shown in Table~\ref{tab:data}, to a steeper and deeper Higgs potential, while at the same time the gauge coupling gets weaker. This is different from the trajectories investigated, which started from the Higgs-like domain~\cite{Maas:2026dlq}, where this stayed basically constant.

\begin{table}
\centering
\begin{tabular}{| c || c | c || c | c | c |}
    \hline
    $\kappa_F$ & $av$ & $am_{\Psi}$ & $g$ & $\alpha_W$ & $\tilde{\lambda}$ \\
    \hline\hline
    $\, 0.12 \,$ & $\, 0.2131(2) \,$ & $\, 0.7933(7) \,$ & $\, 1.91 \,$ & $\, 0.290 \,$ & $\, 4.91 \,$ \\
    \hline
    $\, 0.13 \,$ & $\, 0.23141(13) \,$ & $\, 0.5444(3) \,$ & $\, 1.75 \,$ & $\, 0.243 \,$ & $\, 6.66 \,$ \\
    \hline
    $\, 0.14 \,$ & $\, 0.25196(10) \,$ & $\, 0.26413(12) \,$ & $\, 1.54 \,$ & $\, 0.189 \,$ & $\, 7.67 \,$ \\
    \hline
\end{tabular}
\caption{The vacuum expectation value and the fermion mass in lattice units with statistical errors only in the Higgs-like domain. Additionally, values for the tree-level gauge coupling~$g$ (or~$\alpha_W$) and the four-Higgs coupling~$\tilde{\lambda}$, defined in~\cite{Maas:2026dlq}, are provided. These values allow for a comparison with the Standard Model parameters,~$g_{\text{SM}} = 0.32$ (corresponding to~$\alpha_W = g_{\text{SM}}^2/(4\pi) = 0.0081$) and~$\tilde{\lambda}_{\text{SM}} = 0.25$.}
\label{tab:data}
\end{table}

The transition region does not signal any actual phase transition, but rather a cross-over-like behavior, at least within our resolution\footnote{There may be still a transition associated with a non-local order parameter, like in the bosonic case~\cite{Greensite:2017ajx,Greensite:2018mhh}. This would have a similar structure like, e.g., chiral symmetry breaking in pure Yang-Mills theory without fermions, and thus be considered the effect of a quenched symmetry.}. In this region also~$m_{0^+}/m_{1^-}$ flips from being smaller than one to greater than one. The same behavior has been observed for the QCD-like-to-Higgs-like transition in the purely bosonic system, and was seen both with and without a phase transition\footnote{The Osterwalder-Seiler-Fradkin-Shenker construction~\cite{Osterwalder:1977pc,Seiler:2015rwa,Fradkin:1978dv} assures in the purely bosonic case that the phase diagram is analytically connected. Still, a phase transition is observed in some regions of the phase diagram~\cite{Maas:2017wzi,Bonati:2009pf}. Current results suggest that in the latter case the ratio indeed jumps from below one to above one, while in case of a crossover it changes smoothly~\cite{Maas:2014pba}.}~\cite{Maas:2017wzi}.

What is also noteworthy is the drop of the scalar mass when approaching the phase transition from the QCD-like region, and the subsequent dependence of the scalar mass in the Higgs-like region, which is not observed for simulations in the pure Higgs-like domain~\cite{Maas:2026dlq}.

Most importantly, the theory does not only have no Yukawa coupling at tree-level, but due to the absence of ungauged fermions, it is actually impossible to generate one radiatively. Thus, the only possibility how fermions can influence the Higgs potential is, in perturbation theory, through (at least) two-loop diagrams, where a gauge boson connects the fermion loop to the Higgs. Perturbatively, such a fermion loop scales like~$p^2\ln m+m^2+\dots$, where~$p$ is the (off-shell) gauge boson momentum. Since the latter is integrated over in the external loop, neither of these terms naturally amplify the effect of the fermions on the Higgs potential towards lower masses perturbatively at fixed order. At least resummation effects need to come into play. Thus, despite the couplings being not too strong in the Higgs-like domain, there is no obvious perturbative mechanism to trigger the transition.

There is also the mysterious mass ratio~$m_\Psi/m_{1^-} = 2$, around which the effect triggers. No threshold opens at this ratio. However, this should be seen in the context that the mass ratio~$m_{0^+}/m_{1^-}=1$, also without an associated threshold opening, is decisive when switching between the Higgs-like and QCD-like domain. This pattern is also maintained here. In fact, this feature has been universally observed in any lattice simulations without extended custodial symmetry~\cite{Maas:2017wzi}, i.e., a custodial group larger than the gauge group\footnote{If it is larger, (partially) spontaneously breaking it can create physical Goldstone bosons, which are lighter than the vector particles. Such is the case in multi-Higgs doublet models~\cite{Lewis:2010ps,Catumba:2025axv}.}. This ratio is significant, as it seems also to control the possibility for a first-order thermal phase transition, which appears to happen only if~$m_{0^+}/m_{1^-}<1$~\cite{Kapusta:2006pm}. But this would be in the QCD-like domain. The presence of leptons does not appear to change this~\cite{Maas:2025abc}, which has far-reaching implications for baryogenesis~\cite{Bambi:2015mba}. There is no apparent mechanism known, which could generate such integer mass ratios to be relevant, without crossing thresholds. This makes the appearance of the triggering by light fermions even more ominous.

While these results show evidence for triggering of the BEH effect by light fermions without any obviously known candidate mechanism, this lattice evidence is not definitive. We thoroughly studied finite-volume effects, see~\cite{Maas:2026dlq}, but no dependence on the line of constant physics and discretization effects, because these were small in similar setups~\cite{Maas:2014pba,Maas:2026dlq}. The nature of the transition will require more detailed investigations. These caveats not withstanding, the appearance of this pattern is certainly unusual, and worthwhile to understand further if the conditions under which a BEH effect can occur should be fully understood.

{\bf Conclusions} -- We studied an SU(2)~scalar-fermion-gauge system, starting from the QCD-like domain at infinite fermion mass and lowering it. We encounter a drastic change of the system when crossing the critical mass ratio~$m_\Psi/m_{1^-} =2$, within our precision. The system afterwards exhibits a BEH effect, as all studied observables consistently show. In particular, we see a discontinuous change of the gauge-dependent vacuum expectation value from zero to a finite value, a behavior consistent with a jump. From thereon, when further lowering the fermion mass, the system settles more in the BEH dynamics.

We interpret these observations as an indication that light fermions trigger a BEH effect, offering a new and exciting possibility for its origin. While there is no obvious mechanism at the moment explaining this dynamical phenomenon, it fits well with other observations on critical mass ratios for systems, which can potentially exhibit a BEH effect~\cite{Maas:2017wzi}. This shows that the dynamics of the system are no yet exhaustively understood.

Of course, our fermions are vectorially coupled, and we do not have a Yukawa coupling. To which extent these results can thus be transferred to the Standard Model remains to be seen. A logical next step is to include the Yukawa interactions, and thus also ungauged fermions~\cite{Afferrante:2020fhd}, which are usually assigned a significant role in this context~\cite{Sher:1988mj,Elias-Miro:2011sqh,Alekhin:2012py,Iacobellis:2016eof}. This will move the system closer to the Standard Model. Whether parity violation plays an important role in this context will, however, likely require the use of other non-perturbative methods, due to the current limitations of lattice methods~\cite{Hasenfratz:2007dp}.

{\bf Acknowledgments} -- We would like to thank Antonio Rago for helpful discussions. This research was funded in whole or in part by the Austrian Science Fund~(FWF) [\href{https://www.fwf.ac.at/en/research-radar/10.55776/PAT6443923}{10.55776/PAT6443923}]. The computational results have been achieved using the Austrian Scientific Computing~(ASC5) infrastructure and the Graz Scientific Cluster~(GSC1). The authors acknowledge support from the European Union~(EU) COMETA COST Action~(\href{https://www.cost.eu/actions/CA22130}{CA22130}).

{\bf Data availability statement} -- All data necessary to replicate this analysis and the complete set of ground state energy levels can be downloaded from~\cite{data_release}.

{\bf Code availability statement} -- The version of the \texttt{HiRep} code that was used to produce all configurations and perform all measurements for this paper will be made public in the future.

{\bf Use of generative AI} -- We acknowledge the use of generative AI to support coding. All generated code was independently reviewed, tested, and validated by the authors.

\bibliographystyle{JHEP}
\bibliography{main.bib}

\end{document}